\documentclass[reprint,amsmath,amssymb,amsart,aps,superscriptaddress,twocolumn,float]{revtex4-2}

\usepackage{graphics}
\usepackage{graphicx}
\usepackage{epsfig}
\usepackage{multirow}
\usepackage{dcolumn}
\usepackage{xspace}
\usepackage{array}
\usepackage{tabularx}
\usepackage{gensymb}
\usepackage{bigints}
\usepackage{bm}
\usepackage{subfigure}
\usepackage{color}
\usepackage{mathtools} 
\usepackage{amsmath} 
\usepackage{courier} 

\begin{document}


\title{Coexistent topological and chiral phonons in chiral RhGe: An {\it ab initio} study}
\author{P. V. Sreenivasa Reddy}
\affiliation{Department of Physics, National Taiwan University, Taipei 10617, Taiwan\looseness=-1} 

\author{Guang-Yu Guo}
\email{gyguo@phys.ntu.edu.tw}
\affiliation{Department of Physics, National Taiwan University, Taipei 10617, Taiwan\looseness=-1} 
\affiliation{Physics Division, National Center for Theoretical Sciences, Taipei 10617, Taiwan\looseness=-1}

\date{\today}

\begin{abstract}
The CoSi-family of materials (CoSi, CoGe, RhSi and RhGe) forms a cubic chiral structure and 
hosts unconventional multifold chiral fermions, such as spin-1 and spin-3/2 fermions, 
leading to intriguing phenomena like long Fermi arc surface states and exotic transport properties. 
Recent interest on the phonon behavior in chiral materials is growing 
due to their unique characteristics, including topological phonons, protected surface states 
and the chiral phonons with non-zero angular momentums. 
In this study, we explore the topological and chiral phonon behavior in RhGe, using first-principles 
density functional theory calculations as well as the symmetry and topological analysis.
In particular, we uncover six spin-1 triply degenerate nodal points at the $\Gamma$ point
and six charge-2 double Weyl points at the R point in the Brillouin zone (BZ). Interestingly, 
these topological features are identical to that in the electronic band
structure without the electron spin-orbit coupling, of the same material. 
We expect that this finding not only applies to the CoSi family but also is universal. 
Secondly, we find that chiral crystal RhGe hosts chiral phonon modes with a phonon angular momentum (PAM) and 
an associated phonon magnetic moment (PMM), everywhere in the BZ except at high symmetry points 
such as $\Gamma$, R, X and M. The PAM and PMM are large along the chiral rotation axis
and also in the vicinity of the topological nodes. 
Our study also reveals that all the topological phonon modes are chiral. However, the reverse is not always true. 
Among other things, our finding of the coexistence of topological and chiral phonon modes in chiral RhGe 
not only deepens our understanding of the phonon behavior in the CoSi-family but also opens new pathways 
for developing advanced materials and devices.
\end{abstract}


\maketitle


\section{Introduction}
In condensed matter physics, the electronic topological quantum states lead to the exploration of the quantum spin Hall effect, quantum anomalous Hall effect, Majorana fermions, axion, magnetic monopole and so on \cite{Hasan2010, Qi2011, Kane2005, Konig2008, Chang2013, Fu2008, Hu2020}. These states hold significant importance for practical applications such as quantum computation, thermoelectrics and spin-transfer torques \cite{Lahtinen2017, Xu2017, Fu2020, Mellnik2014}. A key feature of topological electronic states is the presence of unique boundary states guaranteed by bulk-boundary correspondence. These boundary states are resilient to local disorder scattering, making them ideal for low-power electronics and spintronics applications. \cite{Gilbert2021, He2022} 
Beyond electrons, phonons, which are quanta of lattice vibrations and primary heat carriers, play a crucial role in heat conduction, thermal barrier coatings, heat-electricity energy conversion, and superconductivity. The advent of topological electronic states has illuminated new possibilities in the realm of phononics, leading to the development of "topological phononics" \cite{Liu2020, Liu2018, Li2012, Huber2016}. This emerging field leverages quantum concepts like topology, Berry phase, and pseudospin to manipulate phonons in novel ways, potentially revolutionizing applications in phonon waveguides, thermoelectrics, thermal isolation, and other phononic devices \cite{Li2012}. 

Similar to electronic topological states, topological phonons in crystalline materials have been classified 
into several categories: Dirac phonons \cite{Chen2021}, nodal line phonons \cite{Li2020, Zheng2020}, hourglass phonons \cite{Li2021}, 
and Weyl phonons \cite{Miao2018, Li2018, Xia2019}. These topological phonons naturally produce non-trivial, 
topologically protected surface or edge states capable of conducting phonons without scattering. 
This leads to low-dissipation transmission, offering promising properties and a wide range of potential applications \cite{Chen2021a}. 
Unlike electrons but like photons, phonons operate across the entire frequency spectrum without being limited by the Fermi energy 
or the Pauli exclusion principle.~\cite{Chan2018,Wang2023} As a result, the topological phenomena associated with phonons 
are expected to be even richer than those of topological electrons in solid materials.~\cite{Chan2018,Wang2023}

\begin{figure*}
\includegraphics[width=140mm]{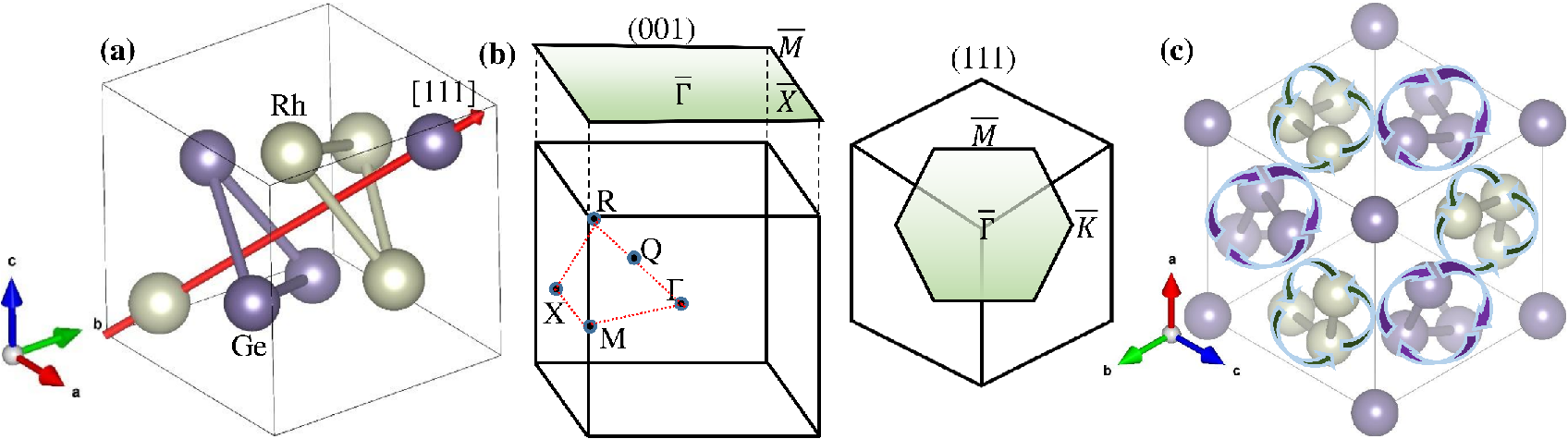}
\caption{(a) Crystal structure of RhGe in the cubic primitive cell. (b) The bulk BZ and the projected (001) and (111) surface BZ. (c) Top view along the [111] direction [the red line in (a)] of the crystal structure (2×2×2 supercell). Here the transparency of the atoms denotes the depth of the atomic positions from top to bottom. The green and indigo coloured arrows indicate the right-handed and left-handed helicity (chirality) of the Ge and Rh atoms, respectively.}
\end{figure*}

Phonons with pseudoangular momentum \cite{Zhang2015} or angular momentum \cite{Zhang2014, Garanin2015, Nakane2018} 
are known as chiral phonons. These chiral phonons were studied in several two-dimensional (2D) lattices, 
such as honeycomb lattice \cite{Zhang2015}, Kekul\'{e} lattice \cite{Liu2017}, and kagome lattice \cite{Chen2019}. 
Recently, chiral phonons were also studied in three-dimensional (3D) systems, e.g., elementary tellurium 
(Te) \cite{Zhang2023}, chiral crystal BaPtGe \cite{Zhang2020_1}, dichalcogenides \cite{Zhang2020}, 
multiferroics \cite{Juraschek2017, Juraschek2019}, magnetic topological insulators \cite{Kobialka2022}, 
CoSn-like kagome metals \cite{Ptok2021}, and binary compounds ABi (A = K, Rb, Cs) \cite{Skorka2023}. In this context, it is worth mentioning some systems containing chiral chains, like $\alpha$-SiO$_2$ \cite{Chen2022}, $\alpha$-HgS \cite{Ishito2023}, binary compounds ABi (A = K, Rb, Cs) \cite{Skorka2023}, or nonsymmorphic systems \cite{Zhang2022}. Chiral phonons can also be observed in a system under strain \cite{Rostami2022}. 
Phonon angular momentum (PAM) and the underlying forces that are responsible for its microscopic origin play a crucial role in the diverse range of effects ranging from the phonon Hall effect \cite{Strohm2005, Grissonnanche2020, Park2020, Flebus2023}, magnetic moment of a phonon \cite{Park2020, Juraschek2017, Juraschek2019, Ren2021, Xiong2022}, Einstein de Haas effect \cite{Zhang2014, Garanin2015, Nakane2018, Tauchert2022}, and topological phononic insulators \cite{Kane2014} to Dirac materials \cite{Hu2021}, driven chiral phonons \cite{Juraschek2020, Geilhufe2021, Geilhufe2023}, and other effects \cite{Hamada2018}.
Regarding the total PAM per unit cell, it is known that in systems without spin-phonon coupling or with time-reversal ($T$) symmetry, the total angular momentum is zero \cite{Zhang2014}. However, non-zero values have been reported in systems exhibiting spin-phonon coupling, such as CeF$_3$, Tb$_3$Ga$_5$O$_{12}$, under specific temperature conditions. For instance, in paramagnetic materials CeF$_3$ and Tb$_3$Ga$_5$O$_{12}$, the total PAM per unit cell has been estimated to be 0.02 $\hbar$ and 1.06 $\times$ 10$^{-4}$ $\hbar$, respectively, at temperatures 1.19 K and 5.45 K \cite{Zhang2014}. The connection between phonon topology and phonon chirality is comprehensively discussed in Ref. \cite{Zhang2025}. In particular, the phonon angular momentum is expected to be non-zero in the vicinity of topological nodal points, indicating that topological phonons constitute a special class of chiral phonons.  

When atoms in a solid move along the trajectory of a circularly polarized vibration mode, they trace closed loops, 
resulting in the generation of angular momentum. In ionic materials, these circular ion motions induce magnetic moments 
that are connected to the angular momentum through the gyromagnetic ratio of the ions. Due to the differing masses 
of the ions, they travel along distinct orbital radii, producing magnetic moments of varying magnitudes, 
which culminates in a net orbital magnetic moment associated with the phonon mode \cite{Juraschek2019}. 
Orbital magnetic moments of phonons reported \cite{Juraschek2019} for 35 different materials with different types 
of structures such as rocksolt, wurtzite, zinc-blend, some perovskites materials and monolayer transition metal 
dichalcogenides along with phonon magneton [phonon magnetic moment (PMM) ($\mu_{ph}$)] values at particular infrared (IR) 
active phonon frequencies. These reported phonon magnetic moment values are in the range of 0.002-1.12 $\mu_N$ 
(here nuclear magneton $\mu_N=e\hbar/2m_p$, where $e$ is the elementary charge, $\hbar$ is the reduced Planck constant 
and $m_p$ is the proton rest mass). In the case of quartz ($\alpha$-SiO$_2$) \cite{Ueda2023}, the reported maximum 
phonon magnetic moment value is around 0.08 $\mu_N$ for the phonon modes along Q$_1$-$\Gamma$ high symmetry direction 
in the BZ. Experimentally, a large effective PMM, approximately 2.7 times to the Bohr magneton ($\mu_B$), 
was observed in Dirac semimetal Cd$_3$As$_2$ \cite{Cheng2020} for the phonon mode at frequency $\sim$ 0.67 THz. 
In the case of Fe$_2$Mo$_3$O$_8$, a polar antiferromagnet material~\cite{Sheu2019}, 
the experimentally observed PMM values are 0.11 $\mu_B$, 
2.0 $\mu_B$ and 2.4 $\mu_B$ for three different modes P1, M1 and M2 at frequecies 42 cm$^{-1}$, 88 cm$^{-1}$ 
and 113 cm$^{-1}$, respectively \cite{Wu2023}. In the case of monolayer MoS$_2$ \cite{Mustafa2025}, 
experimental PMM of $\sim$2.5 $\mu_B$ was observed for doubly degenarate phonon mode E$^{''}$ at frequency 33 meV. 

Coh \cite{Coh2023} recently proposed that, angular momentum in phonons is possible in a crystal if the material presents 
only $P$ (inversion) or only $T$ or only $PT$ or absent of all the three ($P$, $T$ and $PT$) symmetries. 
We have chosen the RhGe material, which belongs to the CoSi family, 
in which the $P$ symmetry is not present, to study the topological and chiral nature of phonons. 
Furthermore, materials of this CoSi family crystallize in a chiral cubic B20 
lattice \cite{Kavich1978,Demchenko2008,Takizawa1988,Engstrom1965,Larchev1982} (See Fig. 1). Interestingly, 
new types of chiral fermions beyond spin-1/2 Weyl fermions, such as spin-3/2 and spin-1 triplet fermions, have recently 
been discovered in chiral crystals including the CoSi family considered here \cite{Bradlyn2016,Chang2017,Tang2017,Chang2018}. 
Unlike spin-1/2 Weyl fermions, spin-3/2 and spin-1 fermionic quasiparticles have no counterpart in high-energy physics, 
and thus are called unconventional (or multifold) chiral fermions. Unlike usual Weyl points, multifold chiral fermion nodes 
sit on high-symmetry points and lines in the Brillouin zone with their chiral charges being larger than $\pm$ 1. 
Furthermore, two partners of a pair of nodal points can be located at two different energy 
levels \cite{Bradlyn2016, Chang2017, Tang2017, Chang2018}. As a result, unconventional chiral fermion semimetals 
were predicted to exhibit exotic physical phenomena such as long Fermi arc surface states \cite{Chang2017, Tang2017, Rao2019}, 
gyrotropic magnetic effect \cite{Zhong2016}, and quantized circular photogalvanic effect \cite{Juan2017}. 
In the CoSi, CoGe, RhSi, and RhGe compounds, two independent nonzero helicity-tunable spin Hall (Nernst) 
conductivity tensor elements were found instead of one element in nonchiral cubic metals \cite{Hsieh2022}. 

In RhGe, alongside its notable electronic structure properties, some phonon studies have been 
conducted.~\cite{Tsvyashchenko2016,Magnitskaya2019, Chtchelkatchev2020,Mardanya2024} 
Measurements of electrical resistivity and magnetization reveal a superconducting state below the transition 
temperature $T_c$ of $\sim$ 4.3 K and weak ferromagnetism below $T_m$ $\sim$ 140 K \cite{Tsvyashchenko2016}. 
Studies on the effects of pressure on the electronic band structure and phonon dispersion of RhGe, up to 43 GPa, 
have shown no phase change; however, electronic topological transitions were observed around 22 GPa, characterized 
by the appearance and disappearance of electron and hole sheets in the Fermi surface \cite{Magnitskaya2019, Chtchelkatchev2020}. 
In this paper, we focuses on exploring the topological and chiral characteristics of the phonons in this material.

The organization of the paper is as follows. Computational details are presented in Sec. II. In Sec. III, we report the crystal structure and phonon dispersion. The topological features in the phonon spectrum are discussed in Sec. IV. Chiral nature of the phonons is discussed in Sec. V. The conclusions drawn from this work are given in Sec. VI.

\begin{table*}
\caption{Character table for the $T_4$, $C_3$ and $C_2$ point groups.}
\begin{tabular} {ccccccccc}
\hline
\hline
$T_4$   & $E $       &$4C_3$        &$4C_3^2$       &$3C_2$ &functions \\
\hline
$A$       &$+1$     &$+1$             &$+1$             &$+1$  & $x^2+y^2+z^2$\\
$E$       &$+1,+1$  &$+\varepsilon,+\varepsilon^*$ &$+\varepsilon^*,+\varepsilon$ & $+1,+1$ &$(2z^2-x^2-y^2, x^2-y^2)$ \\
$T$       &$3$      &$0$      &$0$      &$-1$ &$(x,y,z), (xy,xz,yz), (J_x,J_y,J_z)$\\
\hline
\hline
$C_3$        &$E$      &$C_3$  & $(C_3)^2$ & &functions\\
\hline
$A$       &$+1$     &$+1$     &$+1$  &    &$z,x^2+y^2, z^2, J_z$  \\
$E$ &$+1,+1$ &$+\varepsilon,+\varepsilon^*$ &$+\varepsilon^*,+\varepsilon$ & &$(x, y), (xz, yz), (x^2-y^2,xy), (J_x, J_y)$ \\
\hline
\hline
$C_2$ & $E $ & $C_2$ & &  &functions\\
\hline
$A$    &$+1$     &$+1$ & & & $z,x^2,y^2,z^2,xy,J_z$\\
$B$ &$+1$ &$-1$  & & &$x,y,xz,yz,J_x,J_y$\\
\hline
\hline
$\varepsilon=exp(2i\pi/3)$
\end{tabular}
\end{table*}

\section{Computational details}
In the present study, we perform {\it ab initio} calculations of the total energy, electronic structure and atomic
forces in RhGe, based on the density functional theory (DFT) with the generalized gradient approximation
(GGA) \cite{Perdew1996}. The highly accurate projector-augmented wave (PAW) method \cite{Kresse1999},
as implemented in the Vienna {\it Ab-initio} Simulation Package (VASP) \cite{Kresse1996}, is used.
A large plane-wave basis set cutoff energy of 500 eV is used, ensuring the accuracy of the calculations.
The total electronic energy is converged to within 10$^{-8}$ eV.
A 4 $\times$ 4 $\times$ 4 $\Gamma$-centered k-point mesh is employed for the Brillouin zone integration,
offering a balance between computational efficiency and precision.
The supercell force-constant method, as implemented in the PHONOPY package \cite{Togo2015}, is employed to
calculate the phonon dispersion relations of RhGe.
To determine the interatomic force constants, we use a 3 $\times$ 3 $\times$ 3 supercell, which contains a total of 216 atoms,
providing the sufficient accuracy for convergence.
To explore the topological phonon surface states, including iso-frequency arcs, chirality, and Berry curvature,
we construct a tight-binding Hamiltonian. This Hamiltonian is derived from the calculated force constants
and dynamical matrices, using the Python script \texttt{phonon$_-$hr.py}, which is part of
the WANNIERTOOLS package \cite{Wu2018}. The WANNIERTOOLS package is further employed for detailed topological analysis,
enabling the visualization and computation of topological properties such as phonon surface arcs and Berry curvature.
Additionally, the IR2TB program \cite{Gao2021} facilitates the derivation of irreducible representations
of phonon states from the tight-binding Hamiltonian.

\section{Lattice dynamics}
\subsection{Crystal structure}
RhGe crystallizes in a cubic structure with space group $P2_13$ (space group No. 198) 
as well as lattice constant $a = 4.862$ \AA \xspace \cite{Larchev1982} and Wyckoff position 4a:($u_x$, $u_x$, $u_x$) 
for both Rh and Ge atoms ($x$ = Rh and Ge) with $u_\mathrm{Rh}$=0.135 and $u_\mathrm{Ge}$=0.84. The crystal structure 
of the noncentrosymmetric RhGe is shown in Fig. 1(a). The bulk BZ of the RhGe is shown in Fig. 1(b) 
along with the (001) and (111) surface BZ. RhGe has the tetrahedron ($T_4$) point-group symmetry, 
which provides two twofold screw rotations, $S_{2z} = \{C_{2z}\vert1/2,0,1/2\}$, $S_{2y} = \{C_{2z}\vert0,1/2,1/2\}$, 
and one threefold rotation, $S_3 = \{C_{3,111}^{+} \vert0,0,0\}$, as generators at the $\Gamma$ point. At the R point, 
the generators are $S_{2x}$ = \{$C_{2x}$$\vert$1/2, 3/2, 0\}, $S_{2y}$ = \{$C_{2y}$$\vert$0,3/2,1/2\} 
and $S_3$ = \{$C^{-}_{3,111}$$\vert$0,0,0\}, while at the X point, they are $S_{2y}$ = \{$C_{2y}$$\vert$0, 1/2 , 1/2\} 
and $S_{2z}$ = \{$C_{2z}$$\vert$1/2,0,1/2\} \cite{Cracknell1972}. Because RhGe is nonmagnetic, time-reversal symmetry is present. 
To understand further, we have examined the irreducible representations of the phonon states in RhGe. 
According to the $T_4$ point group character table, 
a total of 12 symmetry operations are present in RhGe with elements $E$, four $C_3$ rotations, four ($C_3$)$^2$, 
and three $C_2$ rotations. The representations are named $A$, $E$, and $T$, where $A$ is singlet, $E$ is doublet, 
and $T$ is triply degenerate. Corresponding character tables for $T_4$, $C_3$, and $C_2$ point groups are provided in Table I.

\subsection{Phonon dispersion} 
We present the calculated phonon dispersion in Fig. 2(a) and the total and atom-projected phonon density of states (DOS) in Fig. 2(b). 
Figure 2(a) shows that there is no imaginary frequency phonon mode, indicating the dynamical stable nature of the RhGe system.
Our calculated phonon spectra agree well with other studies at ambient conditions \cite{Chtchelkatchev2020, Mardanya2024}. 
Previous studies of the electronic structure of the B20 crystals \cite{Hsieh2022,Mardanya2024} showed the significant changes 
in the electronic band structure and topological properties when the relativistic electron spin-orbit coupling (SOC) is included 
in the self-consistent electronic structure calculations. 
To check weather this electron SOC would also affect the lattice vibrations 
or not in the present system, we have performed the phonon dispersion calculations both with and without including the SOC 
in the electronic structure calculations. No discernible change in the phonon dispersion is found when the SOC is included,
as shown in Fig. S1 in the Supplemental Material (SM) \cite{SM}. Thus, for simplicity, we will focus on the results obtained 
without including the SOC in the rest of this paper. 

It is well known that the number of phonon modes is equal to three times the number of atoms in the primitive cell. 
Therefore, there are 24 calculated phonon modes (bands) in Fig. 2(a) because we have eight atoms in a primitive cell.
From the phonon dispersion plot, a strong mixing of acoustic and optical modes is observed 
in all high symmetry directions except along the $\Gamma$-X. Additionally, the phonon modes along the R-M-X-R high-symmetry 
directions exhibit double degeneracy. This arises from symmetry-mediated conditions, specifically the combination 
of twofold screw rotational symmetry and time-reversal symmetry operators, which together form an antiunitary operator. 
This operator enables higher-dimensional band degeneracy in the $k_i = \pi$ plane \cite{Sreeparvathy2022}. 
In contrast, the absence of these symmetry conditions in other high-symmetry directions within the BZ results in 
the lifting of this degeneracy.  

\begin{figure}
\begin{center}
\includegraphics[width=80mm]{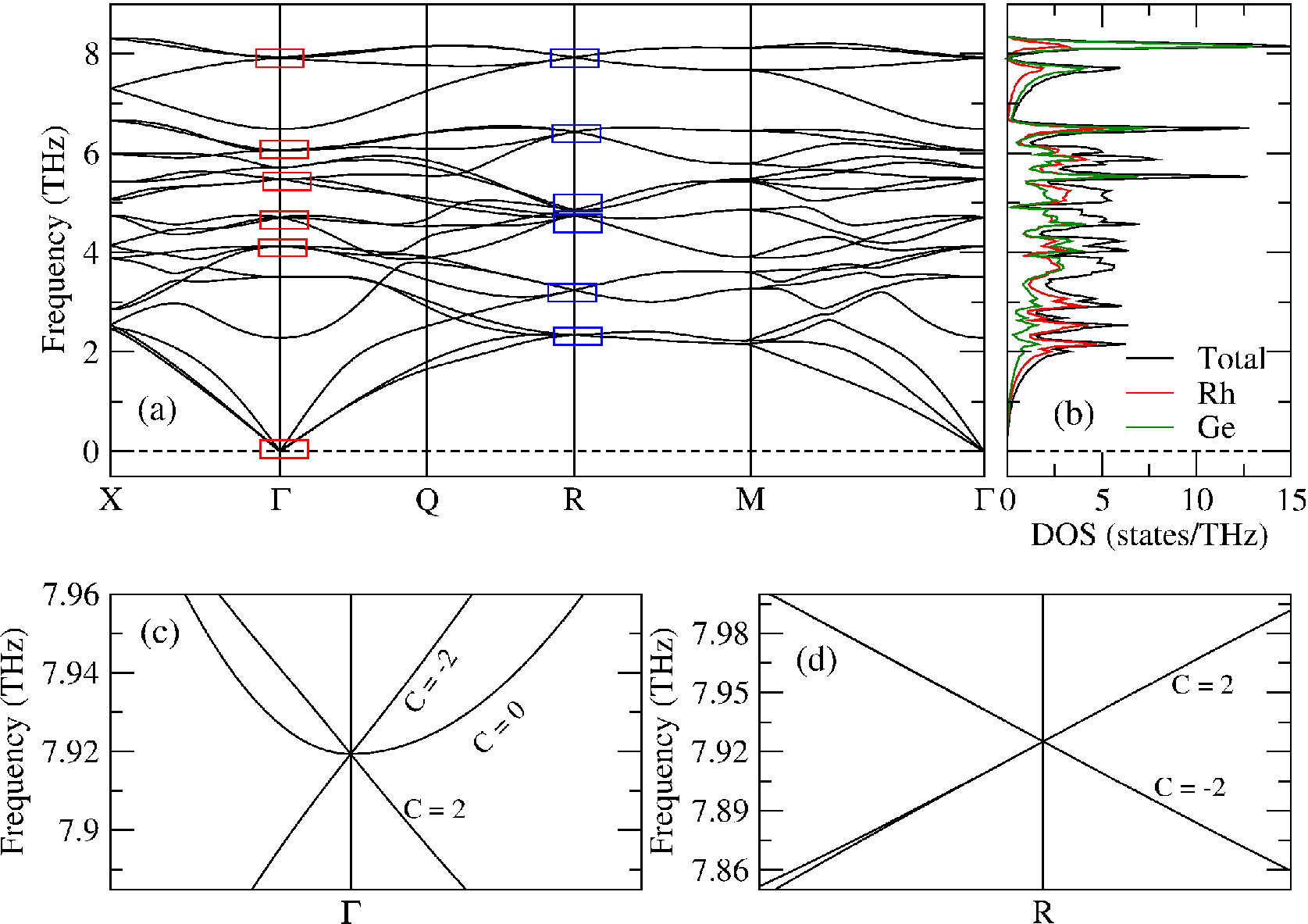}
\caption{(a) Phonon dispersion along the symmetry lines in the BZ and (b) total and atom-projected phonon density of states (DOS) 
of RhGe. The red and blue boxes in (a) denote the locations of the spin-1 nodal points at $\Gamma$ and charge-2 double Weyl points at R, 
respectively. (c) Enlarged plot of the spin-1 nodal point around 7.92 THz at $\Gamma$ and (d) enlarged plot of
charge-2 double Weyl point near 7.92 THz at R. In (c) and (d), $C$ denotes the Chern number of the phonon bands.
}
\end{center}
\end{figure}

At $\Gamma$, the optical modes have different band degeneracies as $\Gamma_{opt}=2A+2E+5T$ where $A$, $E$ 
and $T$ represents, singly, doubly and triply degenerate, respectively. At the R point, 
six fourfold degenerate bands occur. In the phonon spectrum, the upper four optical phonon modes, 
located around the 7 THz frequency region, are distinct from the lower optical modes. 
From the calculated atom projected phonon DOS, modes below 4 THz are dominated by the vibrations 
of the Rh atom. From 4 to 7 THz, the vibrations of both Rh and Ge atoms contribute significantly. Above 7 THz, 
the vibrations from the Ge atom are dominant. 
It is observed that all the modes at $\Gamma$ oscillate to and from along the vector. At other points, for example, 
at {\bf q} = (2$\pi/a$)(0.0025253, 0.0025253, 0.0025253),
some modes have both oscillation and circular motion behavior, which may lead to the chiral nature of the phonons 
in the present system, as will be discussed in section V. 

To compare the phonon behavior of RhGe with other light element-containing members of the CoSi family, we have calculated 
the phonon dispersion and phonon DOS of CoSi, as displayed in Fig. S2 in the SM \cite{SM}. The overall phonon behavior of CoSi 
is very similar to that of RhGe except an increase in the phonon frequencies. This increase in the vibrational 
frequency is due to the smaller masses of the Co and Si elements, as compared with that of the Rh and Ge elements, respectively. 
We thus expect that all other properties like topological and chiral behavior of phonons in the CoSi system will be similar to RhGe.      

\begin{table}
\caption{Infrared (IR) and Raman active modes with 4a Wyckoff position (WP) along with their activity ($\surd$ represents the mode being IR or Raman active).}
\begin{tabular} {cccccccccccccccc}
\hline
\hline
\multicolumn{1}{c}{} & & &  &\multicolumn{4}{c}{IR} & & & & &\multicolumn{4}{c}{Raman} \\
\hline
\multicolumn{1}{c}{WP} & & & & \multicolumn{1}{c}{A}& \multicolumn{1}{c}{$^1$E}& \multicolumn{1}{c}{$^2$E}& \multicolumn{1}{c}{T}  & & & & & \multicolumn{1}{c}{A}& \multicolumn{1}{c}{$^1$E}& \multicolumn{1}{c}{$^2$E}& \multicolumn{1}{c}{T} \\
\hline
\multicolumn{1}{c}{4a} & & & & \multicolumn{1}{c}{.}& \multicolumn{1}{c}{.}& \multicolumn{1}{c}{.}& \multicolumn{1}{c}{3}  & & & & &\multicolumn{1}{c}{1}& \multicolumn{1}{c}{1}& \multicolumn{1}{c}{1}& \multicolumn{1}{c}{3} \\

\hline
\multicolumn{1}{c}{Activity} & & & & \multicolumn{1}{c}{.}& \multicolumn{1}{c}{.}& \multicolumn{1}{c}{.}  & \multicolumn{1}{c}{$\surd$}  & & & & & \multicolumn{1}{c}{$\surd$}& \multicolumn{1}{c}{$\surd$}& \multicolumn{1}{c}{$\surd$}& \multicolumn{1}{c}{$\surd$} \\

\hline
\hline
\end{tabular}
\end{table}

\subsection{Infrared and Raman active modes in RhGe}
In the present system, both Rh and Ge atoms reside at 4a Wyckoff positions. In terms of the irreducible representations (IRREPS) 
of the $P2_13$ space group, the mechanical representation \cite{Kroumova2003} of the modes at $\Gamma$ can be written 
as $M= 2A+2^1E+2^2E+6T$, which are tabulated in Table II where infrared (IR) active and Raman active modes are indicated.  
Table II shows that the present system has $5T$ IR active modes and $2A+2^1E+2^2E+5T$ Raman active modes.
Here only optical modes are considered.
In Table III, we present the details about the IRREPS and activity of each phonon mode at $\Gamma$. 
Table III shows that all the optical modes are Raman active.
We have determined the Raman intensity for the Raman active modes listed in Table III 
using $I_k \sim \sum_{i,j=123} |e^l_iR^k_{ij}e^s_j|^2$. The intensity $I_k$ of Raman line depends on the polarizations 
$e^l_i$ and $e^s_j$ of the incident and scattered light, respectively, and also on the Raman tensor element $R_{ij}$. 
This would help in the interpretation of the measured Raman spectra and the correct assignment of the phonons to the symmetry species. 
The Raman tensors for the three symmetry types are given in Table IV. 
We also evaluate the Raman intensity peaks at frequencies 2.2725 THz, 5.718 THz, 6.063 THz, 6.492 THz, 7.938 THz 
(see supplimentary note C and Fig. S3 in the SM \cite{SM}). 
These frequencies can be attributed to mode numbers 4, 16-17, 18-20, 21, 22-24, which are Raman active and are 
identified through the irreducible representations (IRREPS) (see Table III). 

\begin{table*}
\caption{Mode frequencies ($\omega$), irreducible representations (IRREPS) and infrared (IR)
and Raman (R) activities at the $\Gamma$ point in RhGe.}
\begin{tabular} {ccccccccccccc}
\hline
\hline
 Modes &1-3 &4 &5 &6 &7-9 &10-12 &13-15 &16 &17 &18-20 &21 &22-24 \\
\hline
$\omega$ (THz) &0.0008 &2.281 &3.508 &3.508 &4.117 &4.702 &5.476 &5.711 &5.711 &6.056 &6.482 &7.919 \\
\hline
IRREPS &T &A &$^1$E &$^2$E &T &T &T &$^1$E &$^2$E &T &A &T \\
\hline
Activity &IR &R &R &R &IR+R &IR+R &IR+R &R &R &IR+R &R &IR+R \\
\hline
\hline
\end{tabular}
\end{table*}

\begin{table*}
\caption{Raman tensors for the three symmetry types in the $T_4$ point group.}
\begin{tabular} {cccccccccccccccccccccccc}
\hline
\hline
$A: \begin{pmatrix}  a & 0 & 0 \\ 0 & a & 0 \\ 0 & 0 & a \end{pmatrix}$ &
$E: \begin{pmatrix}  b+c\sqrt{3} & 0 & 0 \\ 0 & b-c\sqrt{3} & 0 \\ 0 & 0 & -2b \end{pmatrix}$, &
$\begin{pmatrix}  b-c\sqrt{3} & 0 & 0 \\ 0 & c+b\sqrt{3} & 0 \\ 0 & 0 & -2c \end{pmatrix}$ &
$T: \begin{pmatrix}  0 & 0 & 0 \\ 0 & 0  & d \\ 0 & d & 0 \end{pmatrix}$, &
$\begin{pmatrix}  0 & 0 & d \\ 0 & 0 & 0 \\ d & 0 & 0 \end{pmatrix}$,
$\begin{pmatrix}  0 & d & 0 \\ d & 0 & 0 \\ 0 & 0 & 0 \end{pmatrix}$ \\

 \hline
\hline
\end{tabular}
\end{table*}

\section{Topological features}
In a fermionic band structure, topological nodal points could emerge due to the multifold band 
crossings.~\cite{Chang2017,Tang2017,Bradlyn2016,Zhang2018}
In the CoSi family, in particular, spin-1 threefold nodal points at the $\Gamma$ point and 
charge-2 fourfold double Weyl nodes at the R point have
been found in the electronic band structure calculated without the SOC~\cite{Tang2017,Hsieh2022} 
[See also Fig. S4(a) in the SM~\cite{SM}].
In the relativistic band structure (i.e., with the SOC included),  both the conventional spin-1/2 Weyl 
fermion nodes and unconventional
fourfold spin-3/2 Rarita-Schwinger-Weyl (RSW) nodes would appear at the $\Gamma$ point, 
and also sixfold double spin-1 fermion nodes would
occur at the R point~\cite{Chang2017,Tang2017,Hsieh2022} [See also Fig. S4(b) in the SM~\cite{SM}]. 

We would expect similar topological features in a bosonic band structure.
As discussed in the previous section, there are six threefold degenerate points at the $\Gamma$ point 
and also six fourfold degenerate points at the R point in the calculated phonon band structure, shown in Fig. 2(a).
To study the topological nature of these nodal points, we have calculated the Chern number of these bands,
as displayed in Figs. 2(c) and 2(d). The calculated Chern numbers indicate that all six threefold degenerate points
at the $\Gamma$ point are the spin-1 threefold chiral nodal points [see Fig. 2(c)] and all six fourfold degenerate points
at the R point are charge-2 double Weyl nodes [see Fig. 2(d)]. 
Among them, the lowest frequency spin-1 chiral nodal point at $\Gamma$ is formed by the acoustic phonon modes. 
The Chern number of the longitudinal acoustic phonon mode is zero, but the Chern numbers for the two transverse 
acoustic phonon modes are $\pm2$, thus leading to the spin-1 nature of these phonons. 
When the wavevector $\bf q$ moves away from the  $\Gamma$ point and becomes finite, 
the  mode with zero Chern number (the third phonon band) has atomic vibrations along the $\bf q$ direction
(see Fig. S5 in the SM~\cite{SM}).
This longitudinal mode is thus achiral and has zero angular momentum (see the calculated PAM
along the $q_x$, $q_y$ and $q_z$ directions, respectively, in Figs. S6-S8 in the SM~\cite{SM}). 
In contrast, the remaining acoustic modes with Chern number $\pm2$ have atomic rotations around the $\bf q$-axis,
respectively, with right/left hand helicities (chirality) (see Fig. S5 in the SM~\cite{SM}). 
These two transverse modes are thus
chiral and have finite angular momentum away  from the  $\Gamma$ point (see the calculated PAM
along the $q_x$, $q_y$ and $q_z$ directions, respectively, in Figs. S6-S8 in the SM~\cite{SM}).
Similar behavior was previously reported in FeSi \cite{Zhang2018} and also in Rb$_2$Be$_2$O$_3$ \cite{Jin2021},
which share the same point group as RhGe.  

At the R point, we have six fourfold degenerate phonon modes which have charge-2 double Weyl quasiparticle behavior. 
As examples, we plot the phonon dispersion around the spin-1 topological nodal point and charge-2 double Weyl node 
near the $\sim$7.92 THz frequency at the $\Gamma$ and R symmetry points in Fig. 2(c) and 2(d), respectively. 
At the R point, the three screw axes ($C_{2x,2y,2z}$) anticommute with each other and satisfy $C^2_{2i} = -1$ 
which is similar to half-integer spin rotations. This implies that all irreducible representations of the system 
have even dimensions, with the smallest possible representation being two-dimensional. In this two dimensional representation, 
the rotations are expressed by $\pm$$i\sigma_{x,y,z}$. However, since the system is also invariant under time reversal, 
the screw axes must commute with the time reversal operator $T$. This requirement dictates that all matrix representations 
of the three screw axes must be real. For two-dimensional $SU(2)$ representations, having real matrices is impossible. 
Nevertheless, it can be achieved with four-dimensional representations. Finally, because the $T$-symmetry preserves 
the Chern number of a topological nodal point, the four-dimensional representation must have a charge of $\pm2$. 
At the R point, which is located at the BZ corner, all the bands form charge-2  double Weyl points. 
According to no-go theorem \cite{Yu2019}, the spin-1 topological nodal point with the Chern number $\pm2$ at $\Gamma$ 
and the charge-2 double Weyl node (with chern number $\mp2$) at R come into pairs. In this material, thus, the phonon bands 
at the  $\Gamma$ point are connected to the fourfold-degenerate double Weyl point at the R point. 
This behavior arises from the combined effects of the $T$-symmetry and the nonsymmorphic nature of the space group, 
which includes three screw axes. 

\begin{figure*}
\includegraphics[width=110mm]{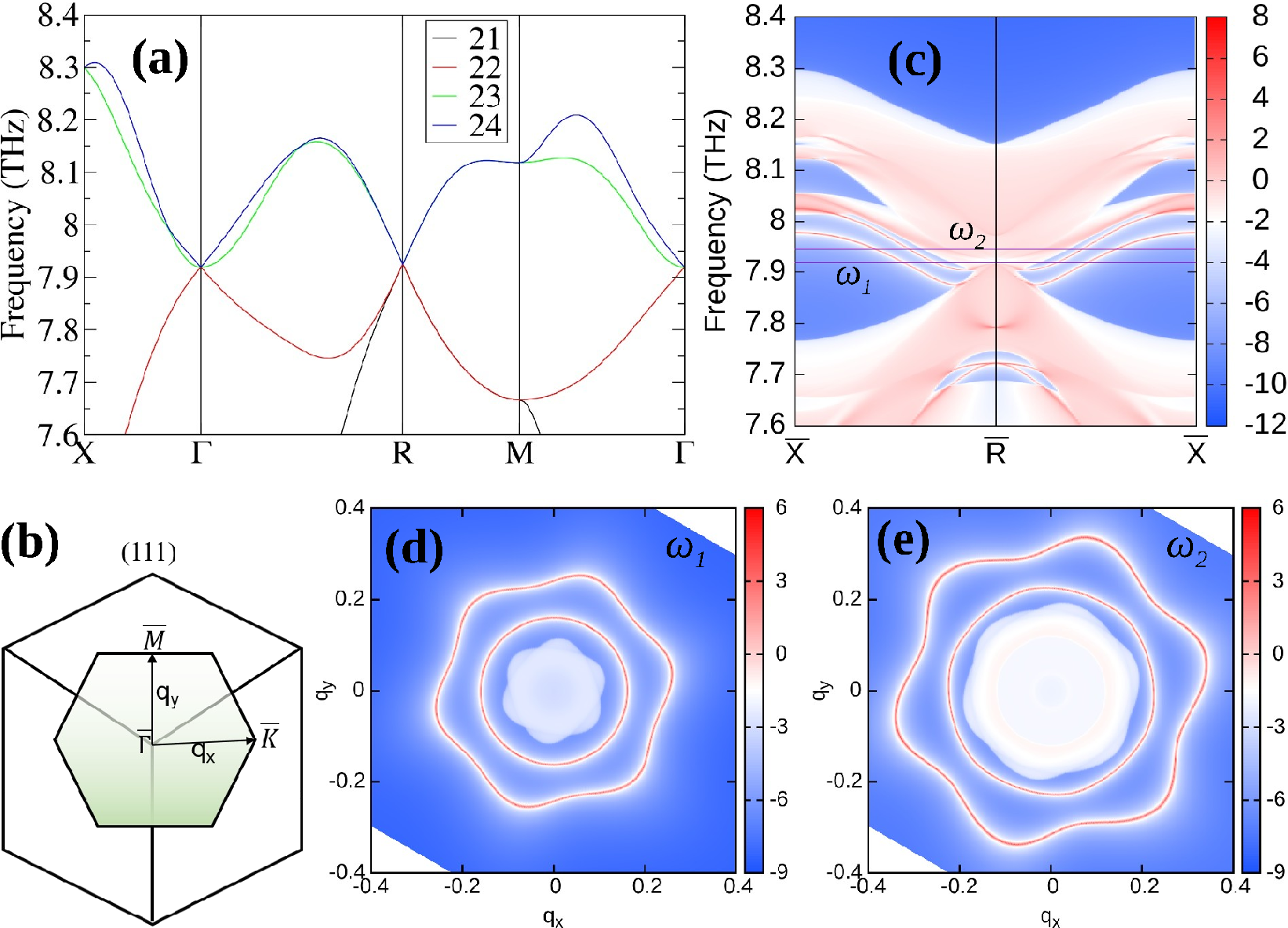}
\caption{(a) Bulk phonon dispersion, (b) (111) surface BZ, and (c) (111) surface phonon dispersion of RhGe. In (c), $\omega_1$ and $\omega_2$ 
are two different frequencies near the double Weyl point. (d) and (e) Surface arcs at $\omega_1$ and $\omega_2$ frequencies, respectively.}
\end{figure*}

\begin{figure*}
\includegraphics[width=110mm]{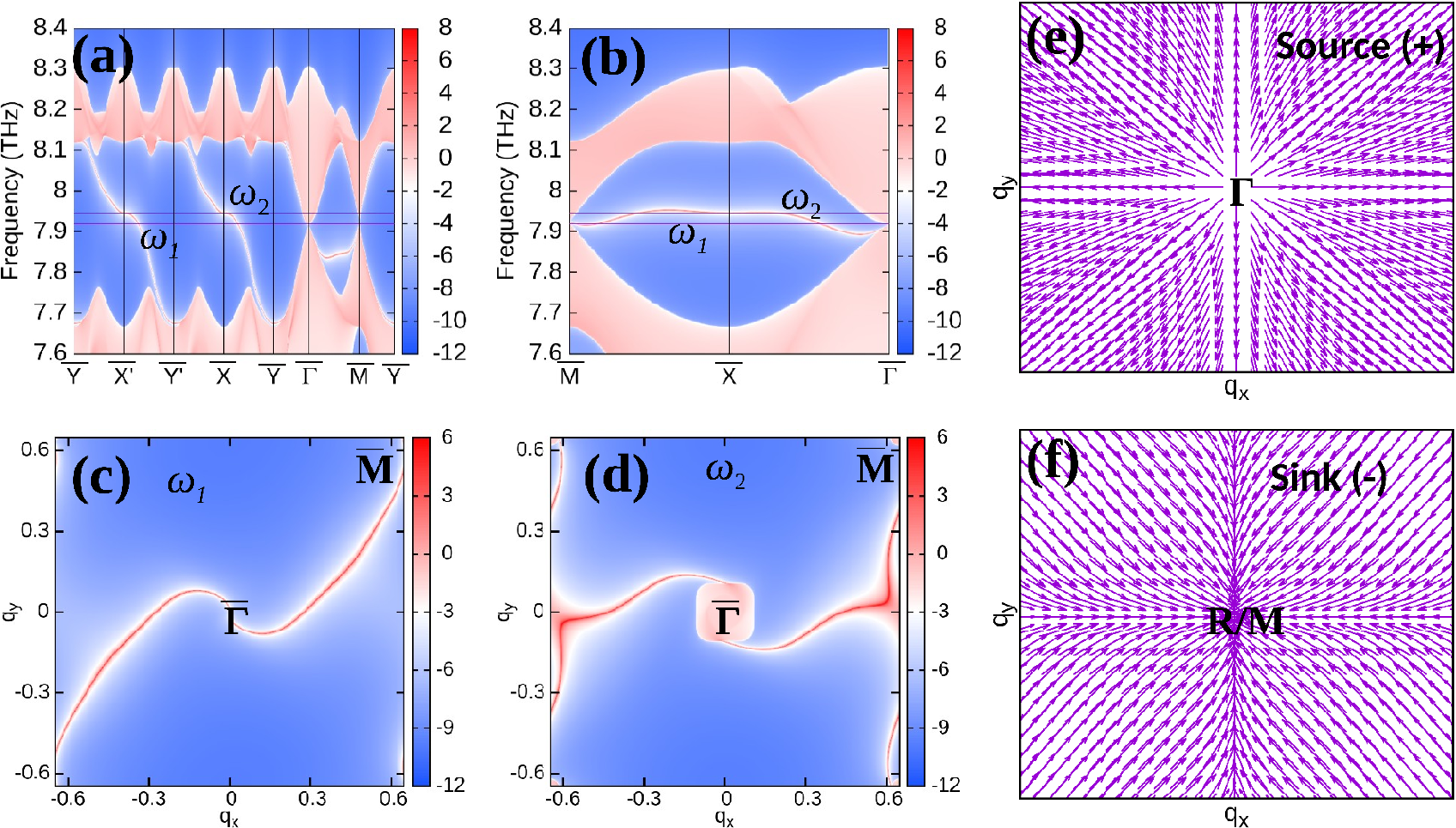}
\caption{(a) and (b) (001) surface phonon bands of RhGe along symmetry lines in the (001) surface BZ [see Fig. 1(b)].
Here $\omega_1$ and $\omega_2$ are two different frequencies near the double Weyl point. (c) and (d) 
Surface arcs at $\omega_1$ and $\omega_2$ frequencies, respectively. (e) and (f) Berry curvatures
around the $\Gamma$ (source) and R/M (sink) points, respectively. }
\end{figure*}

To gain further insight into the topological nontrivial nature of the topological phonons, we have calculated the 
related (111) and (001) surface states of the highest frequency topological phonons from 7.6 THz to 8.6 THz, as plotted in Fig. 3(a)
and Figs. 4(a) and 4(b), respectively. The calculated surface local density of states (LDOS) for this frequency range is plotted 
in Fig. 3(c) along high symmetry directions in the (111) surface BZ [Fig. 3(b)]. The analysis of the (111) surface states 
indicates the presence of charge-2 double Weyl nodes. The corresponding iso-frequency contours are displayed in Figs. 3(d) and 3(e),
corresponding to frequencies $\omega_1$ and $\omega_2$, respectively, indicated in Fig. 3(c). In a similar way, 
the calculated surface LDOS for the same frequency range are plotted in Figs. 4(a) and 4(b) along high symmetry directions 
in the (001) surface BZ. The corresponding iso-frequency contours are plotted in Figs. 4(c) and 4(d), respectively,
corresponding to frequencies $\omega_1$ and $\omega_2$, indicated in Figs. 4(a) and 4(b). The surface state corresponding 
to the bulk BZ spin-1 nodal point and charge-2 double Weyl point projected on the $\Gamma$ and M points on the (001) surface BZ. 
Surface arcs inherently connect two topological nodal points with opposite Chern numbers. Therefore, in the context of double-Weyl points 
located at the center ($\Gamma$) and the corner (M) of the BZ, there should be two distinct arcs connecting these points. 
These arcs are subject to the constraint imposed by the $T$-symmetry, which dictates that they must be symmetric under a $\pi$ rotation 
about the $\Gamma$ point. When considered together, these two arcs form diagonal connections across the BZ, linking the $\Gamma$ and M points 
in a symmetric and balanced manner as evidenced from Fig. 4(c) and 4(d). The calculated Berry curvature at the $\Gamma$ point 
in the $q_x-q_y$ plane is plotted in Fig. 4(e), confirming the flow of the Berry flux from the $\Gamma$ point (source). 
In the similar way, Fig. 4(f) confirms the flow of the Berry flux towards the M point (sink). 

\section{Chiral Phonons}
As discussed already in section III, atoms in RhGe exhibit both oscillatory and rotational behaviors, 
moving along circular trajectories. This results in the formation of closed loops, and hence the vibrational angular momentum. 
In this context, PAM refers to the orbital movement of an atom within a lattice around its equilibrium position. This phenomenon can be likened to the orbital motion of an electron around the nucleus of an atom. 

The PAM can be written \cite{Zhang2014,Coh2023} as, $L^{ph}=\sum_p \sum_j u_j^p \times \dot{u}_j^p$, 
where $u_j^p$ is the displacement vector 
of the $j$th atom in the $p$th unit cell and multiplied by square root mass $m_j$. After expressing the displacement vector 
in the second quantization form and if the system is in equilibrium, the total PAM per unit cell becomes,
 \begin{equation}
 L_{\alpha}^{ph}=\sum_{{\bf q},\nu}[n_{0}(\omega_{{\bf q}\nu})+1/2] l_{{\bf q}\nu}^{\alpha}, \alpha = x,y,z. 
 \end{equation}
The summation is over all the phonon modes $\nu$ and wave vectors ${\bf q}$ within the first BZ. 
 \begin{equation}
 n_0(\omega_{{\bf q}\nu})=[exp(\hbar\omega_{{\bf q}\nu}/k_{B}T)-1]^{-1}, 
 \end{equation}
is the Bose distribution function for the $\nu$th phonon mode 
at wave vector ${\bf q}$ with frequency $\omega_{{\bf q}\nu}$ and $l_{{\bf q}\nu}$ is the mode decomposed PAM, 
 \begin{equation}
 l_{{\bf q}\nu}^{\alpha}=\hbar \epsilon_{{\bf q}\nu}^{\dag} M_{\alpha} \epsilon_{{\bf q}\nu}
 \end{equation}
where $\epsilon_{{\bf q}\nu}$ is a displacement polarization vector or phonon eigenvector for mode $\nu$ at ${\bf q}$, 
and $M_{\alpha}$ is the tensor product of the unit matrix and the generator of $SO(3)$ rotation for a unit cell with $N$ atoms, 
 \begin{equation}
 M_{\alpha}=1_{N \times N} \otimes \begin{pmatrix} 0 & -i\varepsilon_{\alpha\beta\gamma} \\ -i\varepsilon_{\alpha\gamma\beta} & 0 \end{pmatrix}, \alpha,\beta,\gamma \in {x,y,z} 
 \end{equation}
 where $\varepsilon_{\alpha\beta\gamma}$ is the Levi-Civita epsilon tensor. The $x$, $y$ and $z$ components of the PAM can be written as follows
\begin{equation}
l^{\alpha}_{{\bf q}\nu} = \sum_j \sum_{\beta, \gamma} \varepsilon_{\alpha\beta\gamma} \cdot \mathrm{Im} 
\left( \epsilon^{\beta*}_{j, {\bf q}\nu} \, \epsilon^{\gamma}_{j, {\bf q}\nu} \right), \alpha,\beta,\gamma \in {x,y,z},
\end{equation} 
where $\varepsilon^{\gamma}_{j, {\bf q}\nu}$ is the $\gamma$th Cartesian component of the phonon eigenvector for atom $j$.

The magnetic moment of a phonon, which is intrinsically linked to its angular momentum, originates from the circular motion of the Born effective charges.  
This typically results in a small fraction of the nuclear magneton (for detailed formulations, see Refs.  \cite{Zhang2014, Juraschek2017, Juraschek2019, Ueda2023}). 
In the present study, we calculate the atom-resolved phonon magnetic moments ($m_i$) by evaluating the atom-specific Born effective charge tensors ($Z_i^*$) and corresponding gyromagnetic ratios ($\gamma_{i}$). The system under investigation has a cubic symmetry; hence, the Born effective charge tensors are isotropic, i.e., $Z_x^*=Z_y^*=Z_z^*$. The computed Born effective charge values for Rh and Ge atoms are $Z_\mathrm{Rh}^*$ = -4.766 and $Z_\mathrm{Ge}^*$ = 4.766, respectively. The gyromagnetic ratio is defined as $\gamma_{i}= eZ_{i}^{*}/(2M_{i})$, where $e$ is the elementary charge, $Z_i^*$ is the Born effective charge of atom $i$, and $M_i$ is its atomic mass. The index $i$ runs over the atoms in the unit cell. Multiplying $\gamma_{i}$ with Eq. (5) yields the $x$, $y$ and $z$ components of the phonon magnetic moment. 
In the present system, the gyromagnetic ratio values for the Rh and Ge atoms are $\gamma_\mathrm{Rh} = -0.04667 \mu_N/\hbar$ and 
$\gamma_\mathrm{Ge} = 0.06611 \mu_N/\hbar$, respectively. 

For the systems without spin-phonon coupling (or with the time reversal symmetry), the total PAM per unit cell would be zero \cite{Zhang2014}. 
The present system is chiral in nature with the chiral axis along (111). Thus, we have calculated PAM projected 
along the [111]-axis ($J_{111}$).  In Fig. 5, the atom-decomposed and total PAM for all the phonon bands at $T = 0$ K
are displayed along the high symmetry lines in the BZ. Figure 5 shows that both Rh and Ge atoms contribute significantly to the PAM, 
particularly, along the $\Gamma$-Q-R lines in the low to mid frequency range. The color gradients, from deep blue to red, 
correspond to negative and positive angular momentum, indicating the presence of both clockwise and counterclockwise circularly polarized phonon modes. 
Rh atoms show dominant contributions in the lower frequency acoustic modes and several low-lying optical branches, 
whereas Ge atoms exhibit relatively strong angular moment in the mid-frequency optical region. Figure 5 also indicates 
that while some partial cancellation occurs between the Rh and Ge contributions, a net angular momentum survives in some regions, 
with especially strong signals near the avoided crossings and degenerate modes. 
The calculated atom-decomposed and 
total phonon magnetic moment at $T = 0$ K are presented in Fig. 6. Figure 6 reveals patterns similar to those observed in the PAM, 
with the PMM being proportional to the angular momentum weighted by the atomic gyromagnetic ratios. 
Rh atoms contribute more significantly to the PMM at lower frequencies, whereas Ge atoms dominate in the $\sim$ 4-6 THz range. 

Due to the symmetry constraints, all the angular momentum values vanish at high symmetry points $\Gamma$, R, X and M in the BZ.
Indeed, this is corroborated by our explicit numerical calculations. 
Consequently, to explore the behavior of phonon modes in the vicinity of $\Gamma$, we select ${\bf q} = (2\pi/a)(0.0025253, 0.0025253, 0.0025253)$,
which lies close to the BZ center. The atom-decomposed and total PAM and PMM at this {\bf q}-point at $T = 0$ K and
$T = 300$ K are listed, respectively, in Table S2 and Table S3 in the SM \cite{SM}. 
As discussed in the previous section, RhGe hosts topological phonons, 
which are characterized by non-zero Chern numbers. From Table S2, we observe non-zero PAM associated with these phonon modes. 
For example, among the acoustic modes near $\Gamma$, the first two modes possess non-zero Chern numbers and have finite PAM, 
whereas the third acoustic mode, which has a zero Chern number, shows zero PAM. The non-zero Chern number mandates a non-zero 
PAM around the topological nodal point \cite{Zhang2025}. This means that topological phonons are inherently chiral, 
exhibiting circular atomic motions with a finite angular momentum. However, the reverse is not always true. 
Not all chiral phonons are topological, as chiral behavior can occur almost everywhere in the BZ in
the present chiral crystal (see Fig. 5). 
Interestingly, Table S2 of the SM \cite{SM} shows that modes 1 and 2 exhibit the largest PAM and also PMM among all phonon modes,
although the PMM values have opposite signs. These correspond to the acoustic branches near the $\Gamma$ point. 
The PAM values of these modes reache up to 0.5 $\hbar$, i.e., the maximum value that a phonon mode can have at $T = 0$ K. 
The atom-resolved PMM values are about 0.013 $\mu_N$, with Rh and Ge atoms contributing nearly equal 
and opposite moments. As a result, the total magnetic moment for these modes almost vanishes due to cancellation. 
Modes at 5.473, 5.479, 6.055 and 6.057 THz frequencies also show rather large PAM values in the range of $\sim$ 0.2-0.25 $\hbar$. 
These modes are both infrared and Raman active (see Table III), as revealed by the symmetry analysis, 
suggesting strong coupling to external fields. 

To investigate the temperature dependence of PAM and PMM, we have also computed the PAM and PMM values at $T =300$ K,
as listed in Table S3 in the SM \cite{SM}. This temperature is sufficient to thermally populate all phonon modes 
across the entire vibrational spectrum of RhGe, ensuring that contributions from all phonon frequencies are included in our analysis. 
Furthermore, conducting measurements at 300 K is experimentally feasible and would not pose significant challenges. 
At $T = 300$ K, the Bose occupation significantly enhances the phonon population, especially for low-frequency modes, 
resulting in a substantial increase in both PAM and PMM values compared to the zero-temperature case. 
Table S3 indicates that the first mode at 0.012 THz shows an extraordinarily high PAM of approximately 519 $\hbar$, 
arising from the large Bose factor associated with this low-frequency mode. This thermal amplification also leads 
to pronounced contributions from the Rh and Ge atoms, which nearly cancel each other in the total PMM due to their opposite gyromagnetic ratios. 
In contrast, mode 2 still exhibits significant values but to a lesser extent, reflecting the reduced Bose occupation 
with increasing frequency. At higher frequencies (above 4 THz), the Bose factors decrease rapidly, leading to a decay 
in both PAM and PMM magnitudes. Nevertheless, several modes remain magnetically and dynamically active 
due to their symmetry properties. Notably, modes 10 and 12 show large but opposite contributions to the PAM and PMM, 
implying strong dynamic motion that may couple to the applied optical fields, especially given their IR and Raman activity. 
Furthermore, mode pairs such as (13, 15) and (18, 20) again show symmetry-related behaviors with reversed angular momentum 
and PMM directions, consistent with time-reversal pairings. The mode at 5.479 THz (mode 15) shows a substantial positive PAM and PMM, 
while its counterpart (mode 13) carries an equally large but negative contribution, reflecting their conjugate character. 
These results emphasize the strong temperature dependence of phonon angular momentum and magnetic moment, 
particularly in low-energy modes, and suggest that thermal activation of phonon chirality could be a viable route 
to enhance magneto-phononic effects in chiral crystals such as RhGe. For comparison, we also list the calculated PAM and PMM 
of RhGe at the Q point in the BZ at zero and 300 K, respectively, in Tables S4 and S5 in the SM \cite{SM}. 
 
\begin{figure*}
\includegraphics[width=170mm]{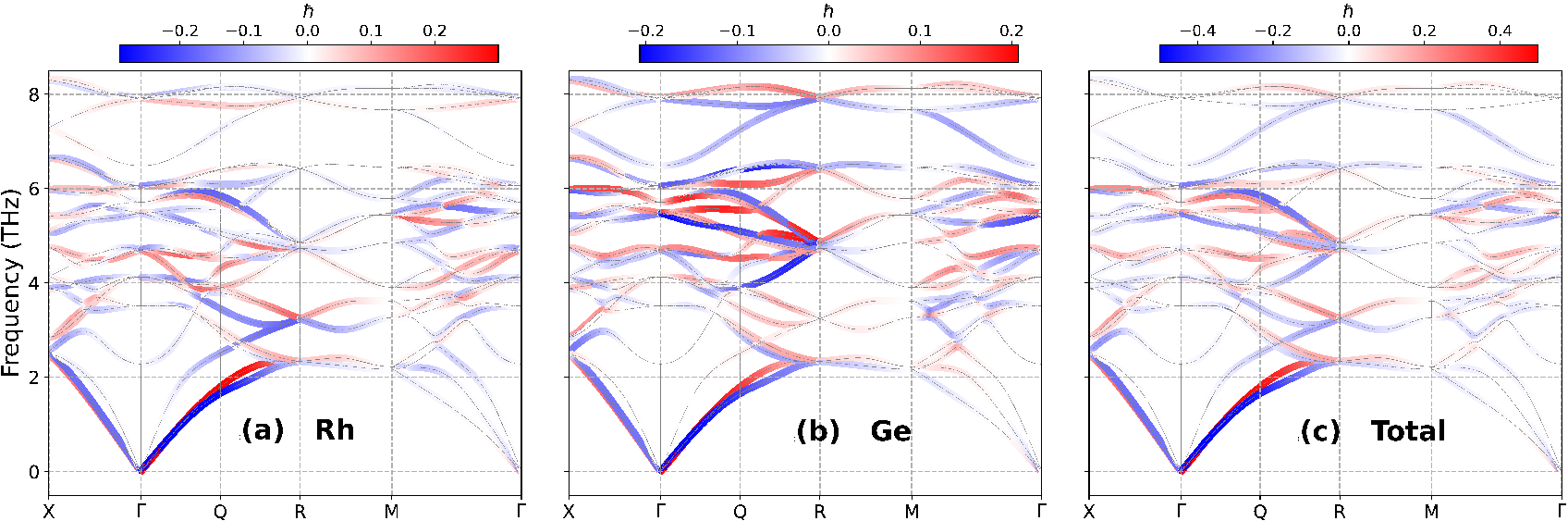}
\caption{(a) Rh atomic, (b) Ge atomic and (c) total phonon angular momentum (PAM) ($J_{111}$) of phonon bands of RhGe
along the symmetry lines in the BZ at $T = 0$ K. Red and blue colors on the color bar indicates positive and negative values of PAM,
respectively. Note that at high symmetry points $\Gamma$, R, X and M,
the PAM values should be zero, as dictated by the symmetry. When plotting the PAM values on the phonon bands,
although the symbols with the minimum size were used, the impression of non-zero values at these symmetry points may be created. }
\end{figure*}

Phonons that exhibit both topological and chiral characteristics open up exciting opportunities across multiple fields. 
In Weyl phonon systems, for instance, chiral fermions can emerge from electronic states near the Fermi surface,
and can enable phonon–fermion coupling \cite{Zhang2025}. This interplay can lead to unconventional responses such as 
non-linear Hall effects and novel spectroscopic signatures. Exploring topological phonon-induced quantum states presents 
a promising research frontier. Furthermore, the established link between pseudospin and Chern number suggests 
that understanding the spin and orbital angular momentum textures of phonons could offer new strategies 
for controlling and manipulating topological and chiral phonons.

\begin{figure*}
\includegraphics[width=170mm]{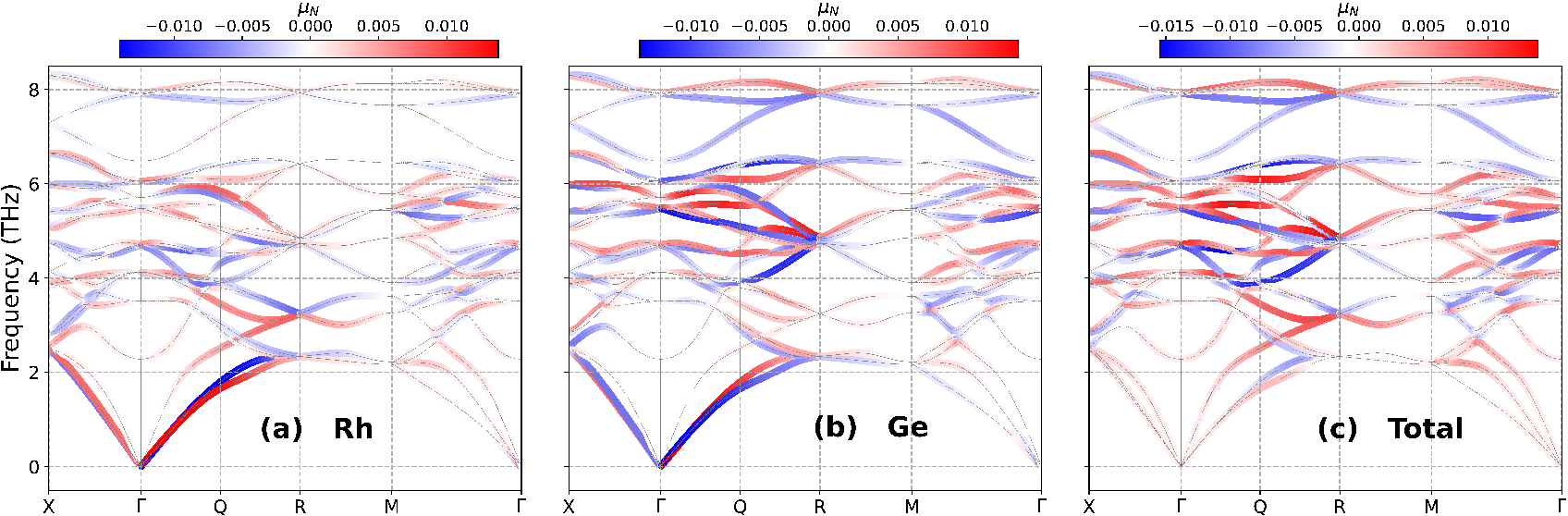}

\caption{(a) Rh atomic, (b) Ge atomic and (c) total phonon magnetic moment (PMM) ($M_{111}$) of phonon bands of RhGe
along the symmetry lines in the BZ at $T = 0$ K. Red and blue colors on the color bar indicates positive and negative values of PMM,
respectively. Note that at high symmetry points $\Gamma$, R, X and M, the PAM values should be zero, as dictated by the symmetry.
When plotting the PMM values on the phonon bands, although the symbols with the minimum size were used, the impression of non-zero values
at these symmetry points may be created.
}
\end{figure*}
  
\section{Conclusions} 
In summary, we have carried out a systematic theoretical study of the lattice dynamics as well as
topology and chirality of the phonon dispersion of noncentrosymmetric chiral crystal RhGe,
based first-principles DFT calculations as well as the symmetry and topological analysis of the phonon modes. 
First of all, our study uncovers unconventional topological features, including the presence of multifold 
and multidimensional topological excitations, in the phonon band structure of RhGe.
In particular, six spin-1 triply degenerate nodal points at the $\Gamma$ point
and six chiral charge-2 double Weyl points at the R point are found. Interestingly, these topological features are
identical to that in the electronic band structure without including the SOC, of the same material.
This finding may be attributed to the fact that, unlike electrons, phonons have no relativistic SOC.
We expect that this finding not only applies to the CoSi family but also is universal.
Our calculated phonon surface states and iso-frequency contours on both the (001) and (111) surface Brillouin zones
further confirm the topological nature of the multifold nodal points at the $\Gamma$ and R high symmetry points,
which are protected by the inherent crystalline and time-reversal symmetries of RhGe.

Secondly, we find that chiral crystal RhGe host chiral phonon modes with a finite PAM and an associated finite PMM,
almost everywhere in the BZ except at high symmetry points such as $\Gamma$, R, X and M.
The PAM and PMM are particularly large along the chiral rotation axis (i.e., $\Gamma$-R line) 
and also in the vicinity of the topological nodal points. While the Rh atoms dominate the PAM and PMM at low frequencies, 
the Ge atoms contribute significantly at mid-frequencies (4-6 THz) to the PAM and PMM.
Our study also reveals that all the topological phonon modes are chiral. However, the reverse is not always true.
At finite temperatures, e.g., at room temperature $T = 300$ K, low-frequency phonon modes would exhibit highly 
enhanced PAM and PMM due to greatly increased Bose occupation caused by thermal excitation.

We expect that the predicted topological and chiral phonons can be detected experimentally by such means as 
inelastic X-ray scattering, Raman and resonant inelastic X-ray scatterings \cite{Miao2018, Ishito2023, Zhang2023, Ueda2023}. 
In particular, the PAM could be observed experimentally by using the tetrahertz sources to excite the phonons 
resulting in large vibrational amplitudes. Thus, the effect of the PAM become visible on a macroscopic level like by 
the interaction of the magnetic or valley degrees of freedom of a material \cite{Nova2017, Shin2018}. 
We also expect that the phonon magnetic moments of RhGe would be observable with experimental techniques 
like helicity-resolved magneto-Raman spectroscopy measurement \cite{Mustafa2025}, and Farady rotation measurements, 
which are able to detect changes in the electronic magnetic order \cite{Nova2017} and nitrogen-vacency 
center magetometry \cite{Casola2018, Taylor2008, Maze2008, Degen2008}. 
Our findings open exciting opportunities for controlling thermal and magnetic properties at the nanoscale, 
with potential applications in thermal management, spintronics, and quantum technologies. 
The realization of non-zero PAM and phonon magnetic moments in RhGe paves the way for phonon-based 
information processing and offers key insights into topological and chiral excitations in complex materials, 
providing a strong foundation for future experimental and technological advancements.

\end{document}